\documentclass[a4paper]{article}
\usepackage{amsmath}
\usepackage{graphics}
\usepackage{epsfig} 
\usepackage[colorlinks=true,hypertex]{hyperref}
\hypersetup{urlcolor=blue, citecolor=red}
\usepackage{authblk,fullpage}

\title{Epidemic oscillations: \\ Interaction between delays and seasonality}

\author{G. Abramson\thanks{E-mail: abramson@cab.cnea.gov.ar}}
\affil{Centro At\'{o}mico Bariloche, CONICET and Instituto  Balseiro, 8400 Bariloche, Argentina} 
\author{S. Gon\c{c}alves}
\affil{Instituto de F\'{\i}sica, Universidade Federal do Rio Grande do Sul,Caixa Postal 15051, 90501-970 Porto Alegre RS, Brazil} 
\author{M. F. C. Gomes}
\affil{Laboratory for the Modeling of Biological and Socio-Technical Systems, Northeastern University, 360 Huntington Av., Boston, MA 02115, USA}

\begin{document}
\maketitle

\begin{abstract}
Traditional epidemic models consider that individual processes occur at constant rates. That is, an infected individual has a constant probability per unit time of recovering from infection after contagion. This assumption certainly fails for almost all infectious diseases, in which the infection time usually follows a probability distribution more or less spread around a mean value. We show a general treatment for an SIRS model in which both the infected and the immune phases admit such a description. The general behavior of the system shows transitions between endemic and oscillating situations that could be relevant in many real scenarios. The interaction with the other main source of oscillations, seasonality, is also discussed.\\ This work was presented at the 9th AIMS Conference on Dynamical Systems, Differential Equations and Applications, Orlando, FL, USA, July 1 - 5, 2012. 
\end{abstract}
\bigskip


\section{Introduction}

Many infectious agents are responsible of oscillating epidemics. In some cases these oscillations are almost periodic, appearing and disappearing with time in a well established way. Examples abound, even in classical textbooks~\cite{anderson,murray}, of such measles, typhus or cholera epidemic waves. Some of these diseases have a cyclic natural history, in which a susceptible subject can become infected, then recovered and immune, and finally return to the susceptible state. However, the mere cyclic nature of the disease does not grant an oscillatory behavior of its epidemic, as can be exemplified by gonorrhea~\cite{grassly2005,grenfell2005}. On the other hand, many epidemics are strongly affected by seasonal factors, which could in principle produce a yearly variation in their prevalence. There are many instances, however, of epidemics that oscillate with \emph{longer} periods: measless~\cite{anderson} and human parainfluenza~\cite{fry2006} every two years, pertussis with a four year period~\cite{anderson}, syphilis with 11 years~\cite{grassly2005}, etc. Even for influenza-like illness, with its apparent yearly period, there exist some indications that its behavior has actually a period of two years~\cite{kuperman}.

How do these oscillations arise in a population, apparently synchronizing the infective state of many individuals? Indeed, it is known that external causes can produce oscillations in epidemic systems: seasonal or El Ni\~no driving in Hantavirus, for example~\cite{abramson2002}. Yet, inherently internal processes are also able to produce oscillations, such as a stochastic dynamics~\cite{risau2007,aparicio2001}, or a complex network of contacts~\cite{kuperman2001}. What are the roles and the interplay of external agents and the dynamic natural history of the disease? 

\begin{figure}[htp]
\centering 
\includegraphics[width=8.4cm, clip=true]{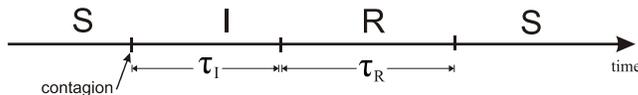}
\caption{Timeline of an individual in an SIRS model, showing the course of the disease after contagion. S, I and R stand for susceptible, infected and recovered respectibly.}
\label{timeline}
\end{figure}

The usual mean-field formulation of epidemic models assumes that the processes that change the states of the individuals proceed at constant rates. For an SIRS system we have three such processes: $S\to I$, susceptibles to infected through contagion, $I\to R$, infected to recovered, and $R\to S$, recovered to susceptible through loss of immunity. So, three rates characterize the dynamics. Let us call $\beta$ the \emph{probability of contagion per unit time} in two-persons interactions. For the two other processes we can use characteristic times: $\tau_I$ is the duration of the infection, the \emph{infectious time} (during which the agent stays in category $I$, infected and infectious), and $\tau_R$ is the \emph{immune time}, the duration of immunity in the state $R$ (see Fig.~\ref{timeline}). The dynamics can be represented as follows: 
\begin{align}
\frac{ds(t)}{dt} &=-\beta\,s(t)\,i(t)   + \frac{r(t)}{\tau_R}, \label{dsdt}\\
\frac{di(t)}{dt} &= \beta\,s(t)\,i(t)   - \frac{i(t)}{\tau_I}, \label{didt}\\
\frac{dr(t)}{dt} &= \frac{i(t)}{\tau_I} - \frac{r(t)}{\tau_R}, \label{drdt}
\end{align}
where $s(t)$, $i(t)$ and $r(t)$ stand for the corresponding fractions
of susceptible, infectious and recovered individuals in the population. Moreover, no demographic change has been included in so $s(t)+i(t)+r(t)=1$ and the system is effectively two-dimensional.

As observed already by Anderson and May~\cite{anderson} this mathematically convenient treatment of the duration of the infection as a constant rate is rarely realistic. It is more common that recovery from infection takes place after some rather well defined time. A similar consideration can be made about the transition from recovered to susceptible. While the ``constant rates'' SIRS model shows, at most, damped oscillations towards an endemic state, it is known (see for example~\cite{hethcote1981}) that sustained oscillations are possible in models with a fixed delay in the recovered to susceptible transition.

In a previous paper~\cite{goncalves2011} we analized an SIRS model with distributed delays both in the infectious and the immune times. In our formulation the fixed times and the constant rates situations are recovered as extreme cases of the probability distribution of the delays. In other words, we can go from the SIRS model with deterministic delays to the classical constant rates SIRS, covering all the intermediate situations in between. 

In the present contribution we analize the interplay of a delayed SIRS model with an external forcing that represents, for example, seasonal variations. We show that a reach dynamics arises, including non-seasonal periodicity of the epidemics of the kind observed in many real scenarios.

\section{Delayed SIRS model}

In \cite{goncalves2011} we developed a general SIRS system in which both the recovery and the loss of immunity processes are described by probability distribution functions of the corresponding times. Let $G(t)$ be the probability (per unit time) of losing infectivity at time $t$ after having become infected at time $t=0$. $G(t)$ can be used as an integration kernel in a delayed equation for the infectious. The individuals that recover from infection at time $t$ are those that got infected at any time $u<t$ and then recover:
\begin{equation}
\beta\int_0^t s(u)i(u)\,G(t-u) du.
\label{kernelG}
\end{equation}
Correspondingly, let us represent the loss of immunity by a second kernel, $H(t)$. Now, the individuals that lose their immunity at time $t$ are those that got infected at earlier times, then recovered at some intermediate time with probability $G$, and finally return to the susceptible state with probability $H$:
\begin{equation}
\beta \int_0^t \left[\int_0^v s(u)i(u)\,G(v-u)du \right] H(t-v) dv.
\label{kernelH}
\end{equation} 
The equations for the evolution of $s(t)$ and $i(t)$ can be written down with the help of expressions (\ref{kernelG}) and (\ref{kernelH}). We have:
\begin{align}
\frac{di(t)}{dt} &= \beta s(t)i(t) - \beta\int_0^t s(u)i(u)\,G(t-u) du, \label{ipunto_delay}\\
\frac{ds(t)}{dt} &= -\beta s(t)i(t)+ \beta \int_0^t \left[\int_0^v s(u)i(u)\,G(v-u)du \right] H(t-v)dv,
\label{spunto_delay}
\end{align}
which, given the constraint $s(t)+i(t)+r(t)=1$, are enough to describe the dynamics.

Appropriate choices of $G(t)$ and $H(t)$ in Eqs. (\ref{ipunto_delay}) and (\ref{spunto_delay}) can reproduce different dynamical behaviors. For example, exponentially decaying distributions with characteristic times $\tau_I$ and $\tau_R$ respectively correspond to the constant rates model (local in time) represented by Eqs. (\ref{dsdt})-(\ref{drdt}). On the other hand, Dirac-delta distributions $G(t)=\delta(t-\tau_I)$ and $H(t)=\delta(t-\tau_R)$ represent a deterministic system in which each of the two processes, recovery and loss of immunity, last exactly $\tau_I$ and $\tau_R$ respectively:
\begin{align}
\frac{ds(t)}{dt} &= -\beta\,s(t)\,i(t) + \beta\,s(t-\tau_0)\,i(t-\tau_0 ), \label{spunto} \\
\frac{di(t)}{dt} &=  \beta\,s(t)\,i(t) - \beta\,s(t-\tau_I)\,i(t-\tau_I ), \label{ipunto} 
\end{align}
where $\tau_0=\tau_I+\tau_R$. Between these two extremes, other suitable $G(t)$ and $H(t)$ are able to represent more realistic situations, in which each process has a maximum at the corresponding characteristic time and some spread around it. One can use a convenient model for these in the form of gamma distributions adequately parametrized. They allow an analytic treatment of the stability of the fixed points. We used:
\begin{align}
G_{k}(t) &= \frac{k^k \,t^{k-1}e^{-kt/\tau_I}}{\tau_I^k \Gamma(k)}, \label{g}\\
H_{h}(t) &= \frac{h^h \,t^{h-1}e^{-ht/\tau_R}}{\tau_R^h \Gamma(h)}. \label{h}
\end{align}
These distributions have mean $\tau_i$ and $\tau_r$, respectively, for any value of the parameters $k$ and $h$. Besides, they cover the range from exponentials (when $k$ or $h=1$) to Dirac delta distributions (when $k$ or $h\to\infty$), with smooth bell-shaped functions for intermediate values of the parameters. 

The model defined by Eqs. (\ref{spunto})-(\ref{ipunto}) must be supplemented with appropriate initial conditions. For an epidemic system, a reasonable choice is an outbreak of infection at the initial time: $i(0)=i_0$, $s(0)=1-i_0$, for example. Due to the nonlocality in time such conditions are insufficient, and extended initial conditions should be provided in order to have a well-posed problem. In more abstract contexts it is customary to provide arbitrary functions $s(t)$ and $i(t)$ in the interval $[-\tau_0,0)$. From an epidemiological point of view, however, it is more reasonable to provide just the initial conditions at $t=0$, and the following complementary dynamics. In the interval $[0,\tau_I)$: no loss of infectivity or immunity, just local contagion (only the first terms in (\ref{spunto})-(\ref{ipunto})). And in $[\tau_I,\tau_0)$: no loss of immunity (absence of the second term in (\ref{spunto})). Equations (\ref{ipunto_delay})-(\ref{spunto_delay}) can also be treated in this way. 

\begin{figure}[tp]
\centering 
\includegraphics[width=8.4cm, clip=true]{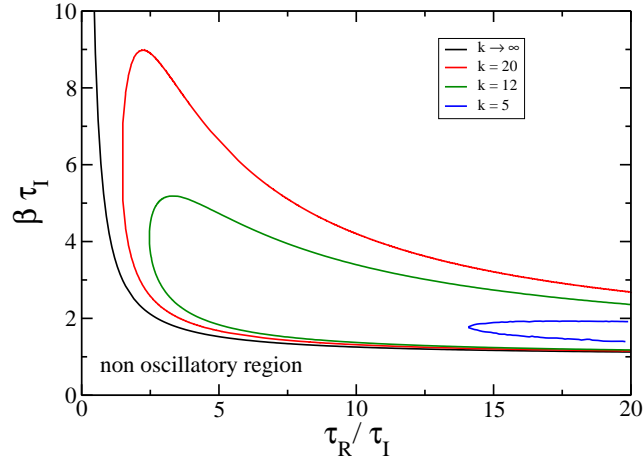}
\caption{Oscillations as controlled by the ratio $\tau_R/\tau_I$ and by the basic reproduction rate $R_0=\beta\tau_I$. Each curve corresponds to a different shape of the kernels that control the recovery and the loss of immunity, $G_k$ and $H_h$ (with $k=h$ for simplicity). The region of oscillations is enclosed by each curve, and above the curve with $k\to\infty$.}
\label{phase-space}
\end{figure}

These nonlocal SIRS epidemic models, at variance with the classic ones, display sustained oscillations. We showed in \cite{goncalves2011}, by means of linear stability analysis and numerical computation of the solutions, that these oscillations appear as Hopf bifurcations in a variety of regions of parameter values. Observe, in Fig.~\ref{phase-space}, the line corresponding to $k\to\infty$, that represents the Hopf bifurcation in the \emph{fixed delays} scenario (where $G$ and $H$ tend to Dirac deltas). The region of oscillations is above this curve. So, there is a transition in the dynamical behavior of the epidemic controlled by the basic reproduction rate of the infection, $R_0=\beta\tau_i$, as is usual in SIR and related models. And there is also a transition controlled by the ratio between the loss off immunity and infectious times, which is controlled by other factors of the disease and not by its reproduction rate (efficacy of vaccines, for example). When the kernels $G$ and $H$ are bell shaped instead of deltas, and there is some indeterminacy in the times of recovery and loss of immunity, the region of oscillations becomes enclosed by the curves shown in Fig.~\ref{phase-space}. There is a reentrance into the region of no oscillations in both directions of the control parameters. When the processes tend to constant rates (the kernels tend to exponentials, when $k\to 1$), the region of oscillations disappears towards the right in the figure. The kernels used in this example have the same shape, with $k=h$, but the situation is the same with differently shaped $G_k$ and $G_h$. Also, numerical solutions and stochastic simulations of the systems concur with these results, that correspond to a linear stability analysis.

\begin{figure}[tp]
\centering 
\includegraphics[width=8.4cm, clip=true]{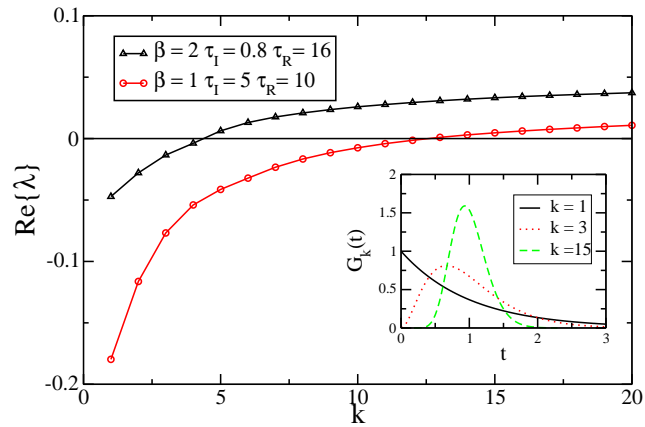}
\caption{Oscillations as controlled by the shape of the probability density functions of the recovery and loss of immunity processes, $G_k$ and $H_h$ (with $k=h$ for simplicity). The main plot shows the real part of the eigenvalue responsible for the destabilization of the fixed point. The inset shows three characteristic shapes of the used kernels, all with mean value 1.}
\label{stability-vs-p}
\end{figure}

The transition from no oscillations at constant rates, to oscillations at fixed delays, also happens as a Hopf bifurcation at a finite value of the shape parameters $k$  and $h$. An example is shown in Fig.~\ref{stability-vs-p}, where we plot the real part of the eigenvalue responsible for the destabilization of the fixed point as a function of the shape parameter $k$. The inset shows three shapes of the kernel corresponding to increasing values of $k$. It can be seen that the oscillations appear when this probability density is still very broad. this behavior represents a further transition in the dynamics of the epidemic system, this one controlled by the \emph{natural history} of the disease.

\section{The role of seasonality}

The external forcing imposed by seasonality can be included in an epidemic model as a periodic oscillation of some of the parameters that control the dynamics. In our present model, defined by $\beta$, $\tau_I$ and $\tau_R$, the sensible choice is a varying contagion rate, $\beta(t)$. In the fixed delays equations this means:
\begin{align}
\frac{ds(t)}{dt} &= -\beta(t)\,s(t)\,i(t) + \beta(t-\tau_0)\,s(t-\tau_0)\,i(t-\tau_0 ), \label{spunto_forced} \\
\frac{di(t)}{dt} &=  \beta(t)\,s(t)\,i(t) - \beta(t-\tau_I)\,s(t-\tau_I)\,i(t-\tau_I ), \label{ipunto_forced} 
\end{align}
with $\beta(t)=\beta_0[1+a\sin(2\pi t/T)]$, oscillating around a mean value $\beta_0$ with an amplitude $a$, and with a period $T$ that could be made equal to 1 year. In the region of oscillations it is expected an interplay of both effects: the oscillation inherent to the delayed dynamics, and that imposed externally through the varying contagion rate. This is, indeed, a situation that mimics real world epidemics in a much more realistic way than the simple model set by Eqs.~(\ref{dsdt})-(\ref{drdt}).

\begin{figure}[tp]
\centering 
\includegraphics[width=8.4cm, clip=true]{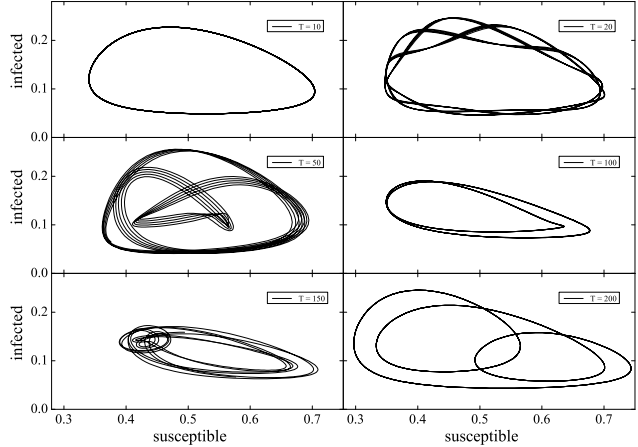}
\caption{Orbits in the $SI$ plane showing oscillating solutions corresponding to the fixed delays model, with $\tau_R=10$, $\tau_I=30$, $\beta_0=0.2$ and $a=0.2$. Each panel shows a different period of the oscillation of the contagion rate, $T$. }
\label{periods}
\end{figure}

Figure~\ref{periods} shows the richness of behaviors that can be expected. The model is in the region of oscillations in the absence of external forcing, and defined by the parameters given in the caption. The plots show orbits (obtained by numerical computation) in the plane of susceptible-infected populations. Each panel corresponds to a different value of the period of oscillation of the contagion parameter $\beta(t)$. Some of the curves are closed and periodic, while others are quasiperiodic or seemingly chaotic. The orbits wind, naturally, due to the nonlocality in time, something that is impossible in the local models due to the uniqueness of solutions. As parameters change these windings change correspondingly, producing the observed range of solutions.

\begin{figure}[tp]
\centering 
\includegraphics[width=8.4cm, clip=true]{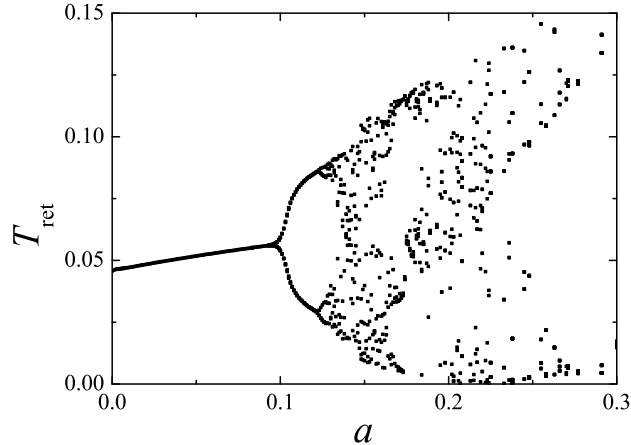}
\caption{Time of recurrence of the outbreaks (maxima of $i(t)$) as a function of the amplitude of oscillation of the contagion rate. The parameters represent an influenza-like illness: $\tau_I=7$, $\tau_R=180$, $T=365$ and $\beta_0=0.18$.}
\label{t-ret}
\end{figure}

\begin{figure}[tp]
\centering 
\includegraphics[width=8.4cm, clip=true]{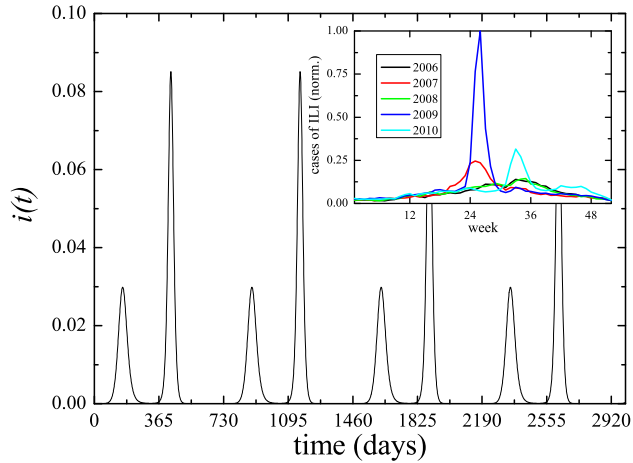}
\caption{Time series of infected corresponding to the period-2 region of Fig.~\ref{t-ret}, with $a=0.12$. Inset: influenza-like illness (ILI, normalized cases) in the South Region of Argentina, plotted as a function of week, for successive five years. }
\label{influenza}
\end{figure}

A compact representation of these transitions can be produced with a kind of Poincar\'e map, representing the time of return to each maximum of $i(t)$. In an epidemic system, this time is the recurrence of the outbreaks of the infection, and is generally well observed in the field. Figure~\ref{t-ret} shows this for a model that represents a situation compatible with influenza (or ILI, influenza-like illness). The characteristic times are chosen as $\tau_I=7$, $\tau_R=180$, $T=365$ which, measured in days, correspond to such an epidemic scenario. $\beta_0=0.18$ is also compatible with influenza, while the amplitude of the oscillation $a$ has been left free and used as a control parameter. A simple recurrence of the epidemic is observed at low values of the amplitude, while a bifurcation cascade leads to more complex behaviors when $a>0.1$. An extended region shows a period-2 behavior, which bears similarity to many cases observed in real systems. The time series of a period-2 solution is shown in Fig.~\ref{influenza}. The successive outbreaks are of different magnitudes. The period is 2 years, as can be seen in the time between two large peaks, while the smaller peaks occur at a different moment of the year. The inset of Fig.~\ref{influenza} shows five consecutive years of cases of influenza-like illness in the South of Argentina \cite{kuperman}, and it is easy to see that even years have their peaks around epidemic week 33 while odd years peak on week 25. The very large outbreak in 2009 corresponds to the epidemic of H1N1 virus (the ``swine flu'').

\section{Conclusions}
We have studied generalized SIRS models with temporally spread infectious and immune processes. These models pose a better description of real epidemic systems than the usual mass-action equations at constant rates, providing a consistent treatment of the time delays inherent of infectious processes. Our model spans a whole range of scenarios in the delay kernels, going from constant rate equations to fixed delays with a single parameter. This allows to model infections that last a more or less fixed time, which is more realistic than constant rates. Even though the mathematical problem is more involved, a stability analysis of the stationary solutions can be made in many relevant cases. Besides this, numerical computation of the solutions, together with stochastic simulations of the processes, provide a reasonable set of tools for the analysis of these systems. 

We have observed that the delayed models show oscillations in their solutions as a function of model parameters, such as the basic reproductive rate and characteristic times. Moreover, oscillations also appear as a function of the width of the characteristic times distributions, which represents the temporal indeterminacy of the corresponding processes. Finally, external forcing representing the seasonal variation of epidemiological parameters can be taken into account. In the case of an oscillating contagion rate, complex oscillations are observed that could be of relevance in the origin of some non-seasonal oscillations observed in real epidemic systems.

%For acknowledgements section, please don't number the section, please begin it with \section*{Acknowledgements}
\section*{Acknowledgments} A cooperative agreement between the Brazilian Coordena\c{c}\~{a}o de Aperfei\c{c}oamento de Pessoal de N\'{i}vel Superior and the Argentinian Ministerio de Ciencia y Tecnolog\'{\i}a (grant CAPES-MINCyT 151/08-017/07) has supported our work. We also thank the Brazilian Conselho Nacional de Desenvolvimento Cient\'{\i}fico e Tecnol\'{o}gico  (grant CNPq PROSUL-490440/2007), the Consejo Nacional de Investigaciones Cient\'{\i}ficas y T\'{e}cnicas (PIP 112-200801-00076), and Universidad Nacional de Cuyo (06/C304).

% You may incorporate your references as follows in your main tex file.
% Using BibTex is not recommended but can be handled.

\end{document}